\documentclass{aastex61}
\usepackage{graphicx,epstopdf,epsfig}

\shortauthors{Nandy et al.}

\begin{document}

\title{Solar Surface Magnetic Field Simulation Enabled Prediction of the Large-Scale Coronal Structure of the 21 August 2017 Great American Eclipse: An Assessment of Model Predictions and Observations}

\author{Dibyendu Nandy}
\correspondingauthor{Dibyendu Nandy}
\email{dnandi@iiserkol.ac.in}
\affiliation{Center of Excellence in Space Sciences India, Indian Institute of Science Education and Research Kolkata, Mohanpur 741246, West Bengal, India}
\affiliation{Department of Physical Sciences, Indian Institute of Science Education and Research Kolkata, Mohanpur 741246, West Bengal, India}

\author{Prantika Bhowmik}
\affiliation{Center of Excellence in Space Sciences India, Indian Institute of Science Education and Research Kolkata, Mohanpur 741246, West Bengal, India}

\author{Anthony R. Yeates}
\affiliation{Department of Mathematical Sciences, Durham University, Durham DH1 3LE, United Kingdom}

\author{Suman Panda}
\affiliation{Center of Excellence in Space Sciences India, Indian Institute of Science Education and Research Kolkata, Mohanpur 741246, West Bengal, India}
\affiliation{Department of Physical Sciences, Indian Institute of Science Education and Research Kolkata, Mohanpur 741246, West Bengal, India}

\author{Rajashik Tarafder}
\affiliation{Center of Excellence in Space Sciences India, Indian Institute of Science Education and Research Kolkata, Mohanpur 741246, West Bengal, India}

\author{Soumyaranjan Dash}
\affiliation{Center of Excellence in Space Sciences India, Indian Institute of Science Education and Research Kolkata, Mohanpur 741246, West Bengal, India}

\begin{abstract}

On 21 August 2017 a total solar eclipse swept across the contiguous United States providing excellent opportunities for diagnostics of the Sun's corona. The Sun's coronal structure is notoriously difficult to observe except during solar eclipses; thus theoretical models must be relied upon for inferring the underlying magnetic structure of the Sun's outer atmosphere. These models are necessary for understanding the role of magnetic fields in the heating of the corona to a million degrees and generation of severe space weather. Here we present a methodology for predicting the structure of the coronal field based on long-term surface flux transport simulations whose output is utilized to extrapolate the coronal magnetic field structures. This prescription was applied to the 21 August 2017 solar eclipse. Post-eclipse analysis shows good agreement between model simulated and observed coronal structures and their locations on the limb. We demonstrate that slow changes in the Sun's surface magnetic field distribution driven by long-term flux emergence and evolution govern large-scale coronal structures with a (plausibly cycle-phase dependent) dynamical memory timescale on the order of few solar rotations -- opening up the possibility of large-scale, global corona predictions at least a month in advance.

\end{abstract}


\section{Introduction}

Solar eclipses have been observed since the ancient times and have been objects of awe and wonder for human beings. While that natural reaction to a rare occurrence of nature has not changed over time, we have come to realize the immense potential for scientific investigations that eclipses provide \citep{2010ApJ...719.1362H}. The solar eclipse that occurred on 21 August 2017 was visible across the contiguous United States. This event, dubbed as the great American solar eclipse generated immense interest among the general public and scientists alike. In particular, this event provided excellent opportunities for detailed observations of the Sun's coronal structure.

The magnetic field configuration of the Sun's atmosphere is responsible for structuring of the million degree corona, its heating and its slow quasi-static evolution. Coronal magnetic field dynamics are also responsible for its brightness and density fluctuations. These dynamics are mediated via magnetic flux emergence and restructuring through relaxation processes and reconnection. The Sun's coronal field is also responsible for determining the heliospheric environment, specifically solar wind conditions, open magnetic flux and cosmic ray flux modulation. Eruptive events such as flares and coronal mass ejections (CMEs) associated with coronal magnetic structures generate severe space weather that may potentially impact human technologies in space and on Earth, including satellite operations, telecommunications, GPS navigational networks and electric power grids \citep{2015AdSpR..55.2745S}. Therefore constraining coronal magnetic field structures is of crucial importance. However, this remains an outstanding challenge even today due to the very low density and consequently low photon flux from the corona (in comparison to the solar disk). In this light, total solar eclipses -- rare events wherein the Sun's bright disk is completely masked by the Moon, provide the best opportunity for ground-based coronal diagnostics.

The Sun's magnetic fields are generated in its interior through a magnetohydrodynamic (MHD) dynamo mechanism \citep{1955ApJ...122..293P,2010LRSP....7....3C} from where they buoyantly emerge to form sunspots. In the Sun's convection zone helical turbulence twists the magnetic field lines at small-scales, and the Coriolis force tilts the axis of underlying magnetic flux tubes at large scales such that bipolar sunspots pairs or solar active regions (ARs) emerge with a tilt. Flux transport processes such as differential rotation, turbulent diffusion and meridional circulation redistribute the magnetic flux driving the surface evolution of the Sun's magnetic fields \citep{1989Sci...245..712W,2011Natur.471...80N}. Surface flux emergence, its consequent redistribution by flux transport processes drive the evolution of the Sun's coronal magnetic field -- which tends to evolve and relax towards a minimum energy state in the low plasma-$\beta$ corona. The latter nonetheless hosts non-potential structures that are far from equilibrium. A variety of theoretical modelling techniques such as Potential Field Source Surface (PFSS) extrapolations, Non-linear Force-Free Field (NLFFF) extrapolations, magneto-frictional and full MHD approaches exist to model the coronal structure. A detailed description of these modeling techniques can be found in \cite{2012LRSP....9....6M}. In the absence of routine, quantitative coronal field observations which are rare (see e.g., \citealt{2017SSRv..210..145C} and references therein) these models are currently relied upon to understand the underlying magnetic field structures that drive coronal dynamics. Efforts are underway to utilize model simulations to generate synthetic polarization profiles for interpreting coronal observations in an effort to marry theory and observations \citep{2016FrASS...3....8G}.

On the one hand, prediction of the coronal field structure during total eclipses utilizing theoretical models is a Rosetta stone for interpretation of the white light corona visible during totality. On the other hand, the observationally inferred coronal magnetic field provides constraints on theoretical modelling efforts. A prediction of the coronal structure of the 21 August 2017 solar eclipse was made based on MHD simulations of the solar corona (utilizing photospheric magnetogram as a bottom boundary condition) \href{http://www.predsci.com/corona/aug2017eclipse/home.php}{Predictive Science Inc.}; see also, \cite{2007ASPC..370..299M,2016ApJ...832..180D}. These simulations provide details of not only the magnetic structure but also the temperature and emission profiles of the corona. However, these simulations are numerically extremely resource heavy, rely on magnetic field maps of only the visible solar-disk (thus missing out on far-side information) and moreover, are not able to capture the impact of built in memory of the solar corona that accrues from the slow large-scale surface field evolution on timescales of months to years \citep{2003SoPh..212..165S}.

Here, we present an alternative modeling approach to large-scale coronal structure predictions and confront this with observations of the 21 August 2017 eclipse. We first utilize a data driven surface flux transport (SFT) model of the Sun (with century-scale solar AR data assimilation) until 16 August 2017 and forward run this to 21 August 2017 (assuming no further emergence of ARs within this period). Subsequently, we use the SFT generated surface field as a bottom boundary condition in a PFSS model to extrapolate the expected coronal fields of the 21 August 2017 solar eclipse. We repeat this procedure for different time-windows of (AR emergence-free) forward runs up to 21 August 2017. We describe our models and data inputs in Section~2, present our results in Section~3 and demonstrate that this method allows predictability of large-scale coronal structures up to a few months in advance. We conclude with a discussion on the implications of our results in Section~4.

\section{Data and Methods}

\subsection{Surface Flux Transport Methodology}
The magnetic field evolution on the solar surface is governed by two physical processes: advection due to the large scale flows and diffusion caused by the turbulent motion of super granular convective cells. The diffusion results in the flux cancellation among the magnetic field of opposite polarities, whereas advection causes transportation of magnetic flux towards the polar regions of the Sun. This entire process is better known as Babcock-Leighton \textit{(BL)} mechanism \citep{1961ApJ...133..572B,1969ApJ...156....1L}. The Surface Flux Transport (SFT) models \citep{1989Sci...245..712W,2000GeoRL..27..621W,1998ApJ...501..866V,2001ApJ...547..475S,2002SoPh..207..291M,2002SoPh..209..287M,2010ApJ...719..264C,2014ApJ...780....5U} are quite successful in capturing the physics of the BL mechanism. For constant diffusivity, the magnetic field evolution on the solar surface is governed by the magnetic induction equation,

\begin{equation}
\frac{\partial \mathbf{B}}{\partial t} ={\nabla} \times (\mathbf{v} \times \mathbf{B}) + \eta \nabla^{2} \mathbf{B}
\end{equation}

\noindent where $\mathbf{v}$ represents the large scale velocities, i.e. meridional circulation and differential rotation present on the solar surface and the parameter $\eta$ symbolizes the magnetic diffusivity. Surface magnetic fields are predominantly radial \citep{1992ApJ...392..310W,1993SSRv...63....1S}. We therefore consider the radial component of the induction equation in spherical polar coordinates given by,

\begin{equation}
\frac{\partial B_r}{\partial t} = - \omega(\theta) \frac{\partial B_r}{\partial \phi} - \frac{1}{R_\odot \sin \theta} \frac{\partial}{\partial \theta}\bigg(v (\theta) B_r \sin \theta \bigg)
+\frac{\eta_h} {R_\odot^2}\bigg[\frac{1}{\sin \theta} \frac{\partial}{\partial \theta}\bigg(\sin \theta \frac{\partial B_r}{\partial \theta}\bigg) + \frac{1}{\sin ^2 \theta }\frac{\partial ^ 2 B_r}{\partial \phi ^2}\bigg] + S(\theta, \phi, t)
\end{equation}

\noindent Here $B_r(\theta, \phi,t)$ is the radial component of magnetic field as a function of the co-latitude ($\theta$) and longitude ($\phi$), $R_\odot$ is the solar radius. The axisymmetric differential rotation and meridional circulation are expressed through $\omega(R_\odot,\theta)\approx\omega(\theta)$ and $v(R_\odot,\theta)\approx v(\theta)$ respectively. The parameter $\eta_h$ is the effective diffusion coefficient and $ S(\theta, \phi, t)$ is the source term describing the emergence of new sunspots. To represent the surface differential rotation as a function of co-latitudes we use an empirical profile \citep{1983ApJ...270..288S}: $\omega(\theta) = 13.38 - 2.30 \cos^{2}\theta - 1.62 \cos^{4}\theta$ (in degrees per day). This profile is validated by helioseismic observations \citep{1998ApJ...505..390S}. For the meridional flow we utilize a profile prescribed by \cite{1998ApJ...501..866V} in our model with a maximum speed of 15 ms$^{-1}$. We have used a constant diffusion coefficient of 250 km$^2$s$^{-1}$ which lies within the values inferred from observations \citep{2000CAS....34.....S}.

\subsection{Sunspot Data}

Modelling the emergence of sunspots requires knowledge of the following parameters: time of appearance, position on the solar surface and the area associated with the spots. In the SFT simulations it is assumed all sunspots appearing on the solar photosphere are Bipolar Active Regions (BMRs) of $\beta$ type and their tilt is based on their latitudinal position \citep{2011A&A...528A..82J}. The Royal Greenwich Observatory (RGO) and United States Air Force (USAF)/National Oceanic and Atmospheric Administration (NOAA) database provides the required data on active regions from August 1913 till September 2016. Using this data, we have calibrated our SFT model by comparing model outputs with observation.

The data required from 1st October 2016 to 16th August 2017 has been obtained from the Helioseismic and Magnetic Imager (HMI) on-board NASA's Solar Dynamic Observatory (SDO). This necessitates cross-instrument calibration \citep{2015ApJ...800...48M}. The area reported by HMI is higher than what is reported by RGO-NOAA/USAF. Since our SFT simulation is calibrated based on 100 years (starting from 1913) of RGO-NOAA/USAF dataset, we scale down the area reported by HMI by a constant factor of $2.2$ to match RGO AR areas. We determine this scaling factor by using a linear fit between the area reported by both databases for the overlapping period of 9 months (January to September 2016). We note that the relative fluxes (of various BMRs and large-scale structures) are preserved in this method. The flux associated with BMRs are estimated based on an empirical relationship \citep{1966ApJ...144..723S,2006GeoRL..33.5102D}: $\Phi(A) = 7.0 \times 10 ^{19} A$ Maxwells, where $A$ is the area of the whole sunspot in unit of micro-hemispheres. This flux is equally distributed among the two polarities of the BMR. The last active region that was inserted in our SFT model is AR 12671 -- which emerged at the surface on 16 August 2017.

\subsection{PFSS Extrapolation}

One of the most common techniques utilized to simulate the corona is the Potential Field Source Surface (PFSS) extrapolation using the photospheric magnetic field as a lower boundary condition \citep{1969SoPh....6..442S, 1969SoPh....9..131A}. This extrapolation assumes a lowest energy (current free) magnetic field corona satisfying the equation
\begin{equation}
\mathbf{\nabla\times B} = 0
\end{equation}
We can therefore assume \textbf{B} to be defined based on a scalar potential ($\Phi$)
\begin{equation}
\mathbf{B} = \mathbf{\nabla}\Phi
\end{equation}
Imposing Gauss's Law we obtain,
\begin{equation}
\nabla^2\Phi=0
\label{eq:finPFSS}
\end{equation}
The PFSS solution is obtained by solving eq. \ref{eq:finPFSS} with a finite-difference method. \cite{1965IAUS...22..202D} argued that magnetic fields outside the Sun must become radial at the point where gas pressure starts dominating over magnetic pressure which is expected to occur at 2.5 $R_\odot$ -- the upper boundary of our PFSS model.

\section{Results}

We use the radial magnetic field data obtained from our SFT simulation to generate the butterfly diagram covering a time from the beginning of solar cycle 23 (May 1996) to 21 August 2017 (Figure~1a). In this simulation, the last observed active region was included on 16 August 2017 (AR 12671) and the model was forward run to generate the predicted surface field distribution on 21 August 2017. The large-scale evolution of the solar surface field driven by AR flux emergence and surface flows is clearly discernable. The last few years of the simulation show the slow poleward migration of ``tongues'' of magnetic flux in the approach to the minimum of cycle 24. It is our premise that the structuring and evolution of the large-scale, global coronal field is due to this surface magnetic field evolution with localized (small-scale) perturbations imposed by the emergence of any new ARs whose effects are only apparent after a certain time -- equivalent to the dynamical memory in the system. The predicted surface magnetic field (Carrington) map centered around 21 August 2017 (the day of the eclipse) is extracted from this simulation and is depicted in Figure~1b (with the solar-disk view depicted in the center-inlay of the first three panels of Figure~2).

This surface magnetic field distribution is then utilized as an input in a PFSS model to extrapolate the global coronal structure expected on 21 August 2017 (Figure~2). The inferred coronal magnetic field structure is our primary and most important prediction and is shown in Figure~2a. Only the open field lines reaching up to the source surface are rendered in Figure~2b. A synthetic coronal white light map reconstructed using an algorithm that provides more weight (i.e., intensity) to positions with a higher density of projected closed field lines (but without any radial gradient of density) is depicted in Figure~2c. This SFT-PFSS coupled model predictions include two prominent helmet streamers (regions of closed field lines whose tops reach the source surface) centered below the equator in the southern solar hemisphere, one each in the East and West limb. Their locations (tip of highest closed loop) are marked as 1 (14$^\circ$ S) and 2 (12$^\circ$ S) in Figure~2a -- their cross-sections appear as voids in Figure~2b, wherein, only open field lines are rendered. A third, more confined and narrow streamer is located in the West limb northern hemisphere (47$^\circ$ N; location 5) whose existence is pinned down in Figure~2b (note the narrow localized void at the same location). This third structure is a pseudo-streamer which separates regions of same-polarity coronal holes \citep{2007ApJ...658.1340W,2015ApJ...803L..12W}. This is borne out in Figure~3a, wherein, location 5 corresponds to a narrow region of negative polarity in the predicted Carrington map separating out two positive polarity patches on the solar surface. Regions marked as 3 and 4, in the northern hemisphere West and East limb, respectively, are latitudinally extended low-lying closed magnetic field structures which merge with the more prominent streamers (1 and 2) in the synthetic white light corona in Figure~2c. The coronal region overlying the northern hemisphere East limb is relatively less active compared to the others. Regions marked as 6 and 7 correspond to open field lines associated with polar coronal holes which appear as unipolar caps in the Carrington maps in Figure~3.

A comparison with the Mauna Loa Solar Observatory (MLSO) coronagraph image acquired on 21 August 2017 (Figure~2d) indicates good agreement between the forward-run model prediction and observed coronal structures. The observations show the existence of two broad helmet streamers, the one on the East limb at 18$^\circ$ S, the one on the West limb at 15$^\circ$ S and the narrow elongated pseudo-streamer on the West limb at 46$^\circ$ N.

To test the underlying physical basis of our prediction methodology, namely that the large-scale global coronal structure is a result of long-term surface magnetic field evolution, and to estimate the memory (time-window) for predictability, we perform model forward-runs to 21 August 2017 without introducing new ARs in the SFT simulation for different time windows (Figure~3). Images in the top panel of Figure~3 show the SFT model predicted large-scale solar surface polarity separations on 21st August 2017. The only difference in these magnetic maps is in the duration of emergence-free evolution. In case of Figure~3a, the last AR is inserted in the SFT simulation 24 days before the eclipse effectively excluding the contributions from two ARs which appeared in this 24 day period. For Figure~3b and Figure 3c, the last AR is inserted 54 days (excluding five ARs) and 76 days (excluding eight ARs) prior to 21 August 2017. Images in the bottom panel show the corresponding PFSS generated coronal open magnetic field lines. A careful study of the surface polarity distributions and coronal magnetic field connectivity shows that the 24 days (AR emergence free) forward-run model predicted large-scale corona is similar to that predicted 5 days in advance. The forward-run for 54 days (i.e., two solar rotations) is qualitatively similar with a slight upward shift in the axis of the West limb pseudo-streamer. The 76 days forward-run predicted corona shows some differences in field line connectivity and latitudinal extent of the (East) limb streamer. We note, however, in spite of these localized differences, the predicted large-scale shape of the corona during the eclipse is not significantly different.  These results indicate a dynamical memory timescale corresponding to two solar rotations in the surface magnetic field evolution induced large-scale solar corona during this (declining) phase of the solar cycle.

\section{Concluding Discussion}

In summary, based on data driven solar surface flux transport simulations forward run to 21 August 2017, the day of the great American solar eclipse, and utilizing the model predicted solar surface magnetic field maps as inputs to a PFSS model, we have generated simulated coronal structures expected to be observed during the eclipse. Post-eclipse comparisons with coronal observations during the eclipse indicate broad agreement between model-predicted structures and their locations on the solar limb. This agreement supports our conceptual idea of utilizing solar surface magnetic flux transport simulation forward-runs to predict the future, large-scale solar corona.

Moreover, through different time-windows of (active region emergence free) model forward-runs and PFSS extrapolations, we demonstrate that a dynamical memory of about one-to-two solar rotations exists in the system for large-scale global coronal structure predictions. This implies that utilizing the methodology outlined herein, it is possible to make continuous forecasts of the global coronal structure expected one month in the future (and plausibly slightly longer with degrading accuracy). This memory is likely to be cycle-phase dependent; during the solar maximum it is expected to be less, and closer to the minimum, it is expected to be more. Future investigations will explore this in detail.

On the one hand, we would like to note that our methodology cannot be applied for predictions of small-scale low-lying coronal structures (due to recent active region emergences) or coronal mass ejections. On the other hand, we would like to emphasize that while the eclipse provides an opportunity to compare the model predicted structures with the observed corona only at the limb, the coupled SFT and PFSS simulations render the global 3-D corona and hence predictions for the large-scale corona across all latitudes and longitudes. These simulations -- being less-resource-intensive -- can therefore be utilized for operational global, large-scale coronal structure forecasting purposes based on the physical principles outlined above. Future studies will explore the possibility of using the predicted coronal field to generate future solar wind and cosmic ray flux modulation to test the usefulness of this technique beyond predictions of the solar corona.

\acknowledgements This eclipse prediction campaign and post-eclipse analysis was conceived and led by CESSI, a multi-institutional Center of Excellence established and funded by the Ministry of Human Resource Development, Government of India. P.B. acknowledges funding by CEFIPRA/IFCPAR through grant 5004-1. S.D. acknowledges funding from the DST-INSPIRE program of the Government of India. D.N. and A.R.Y. acknowledge the NASA Heliophysics Grand Challenge Grant NNX14AO83G for facilitating their interactions. D.N. acknowledges an associateship from the Inter-University Centre for Astronomy and Astrophysics. We acknowledge utilization of data from the NASA/SDO HMI instrument maintained by the HMI team and the Royal Greenwich Observatory/USAF-NOAA active region database compiled by David H. Hathaway. The CESSI prediction of the coronal field expected during the 21 August, 2017 solar eclipse, accompanying images and data are available at the \href{http://www.cessi.in/solareclipse2017/}{CESSI prediction website}. We thank Prosenjit Lahiri and Nabamita Das of CESSI for designing and maintaining the website.

\newpage

\begin{figure}
\gridline{\fig{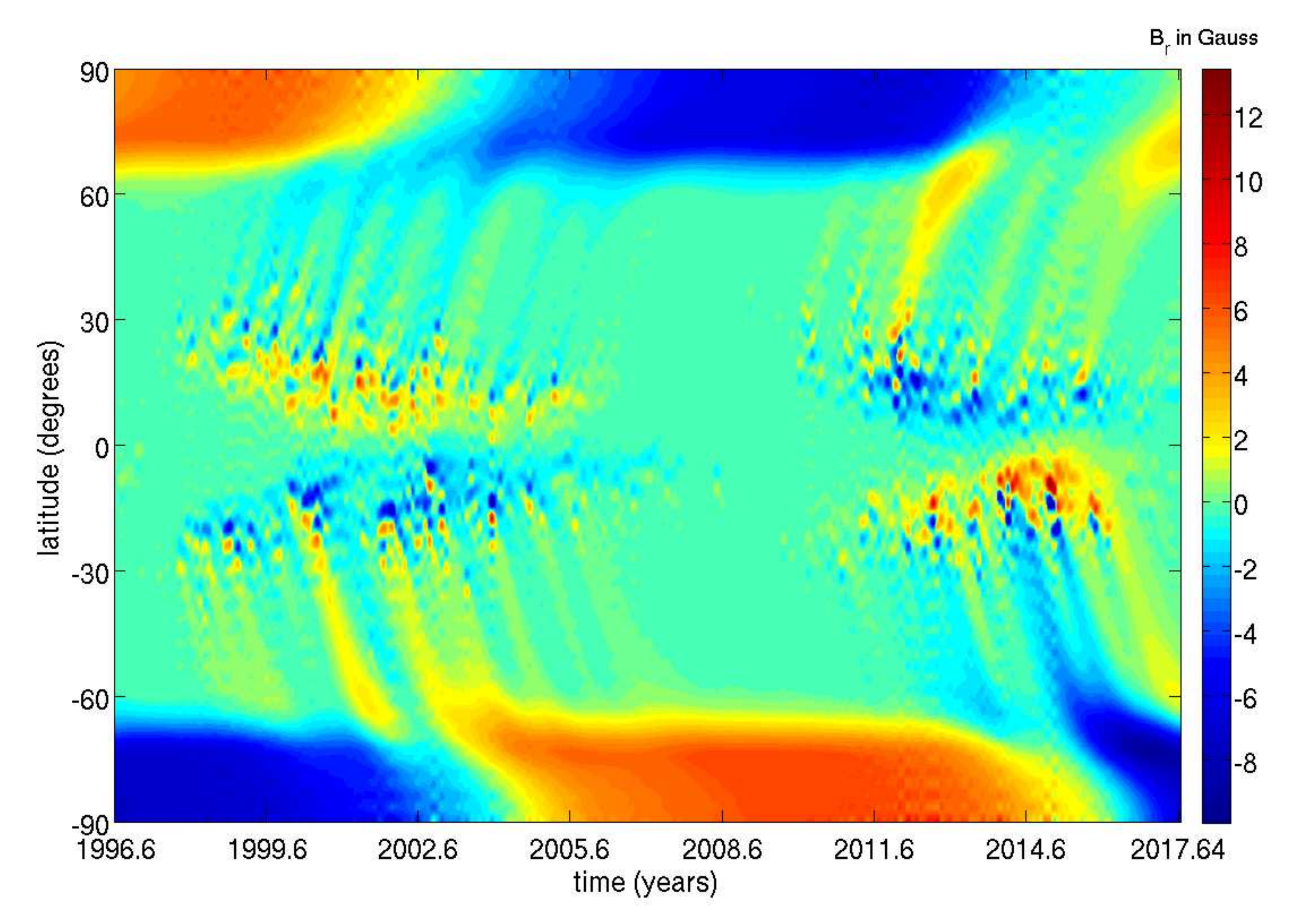}{0.49\textwidth}{(a)}
          \fig{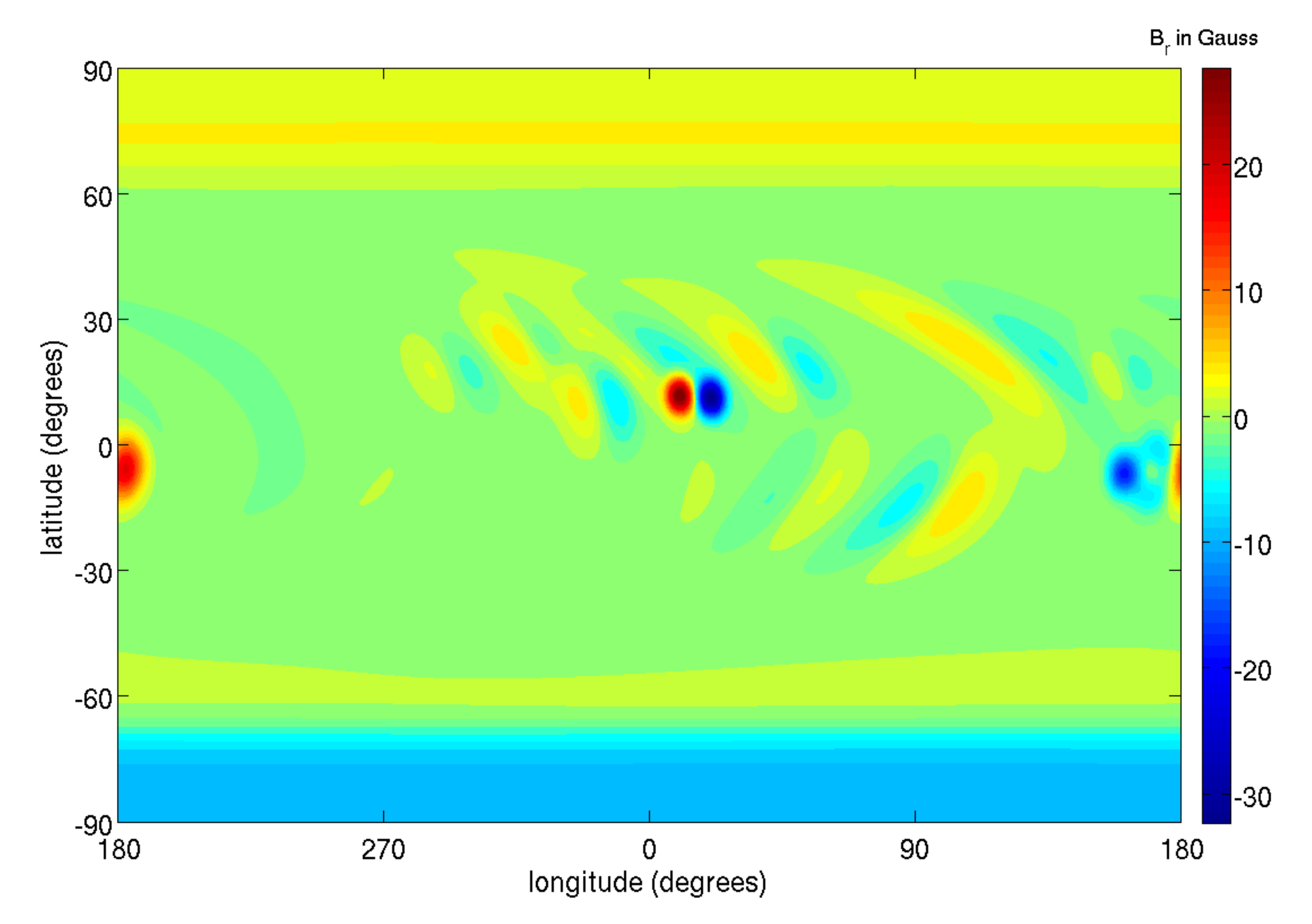}{0.49\textwidth}{(b)}}
\caption{(a): This image depicts the evolution of solar surface magnetic field during the last decade generated from our SFT simulation. The last AR was incorporated on 16 August 2017 and the model was forward run to predict the surface map on 21 August 2017. The radial component of longitudinally averaged magnetic field ($B_{r}$) is plotted as a function of time and latitude. (b): Shows the simulated photospheric magnetic field distribution (Carrington map) of 21 August 2017.}
\end{figure}

\begin{figure}
\gridline{\fig{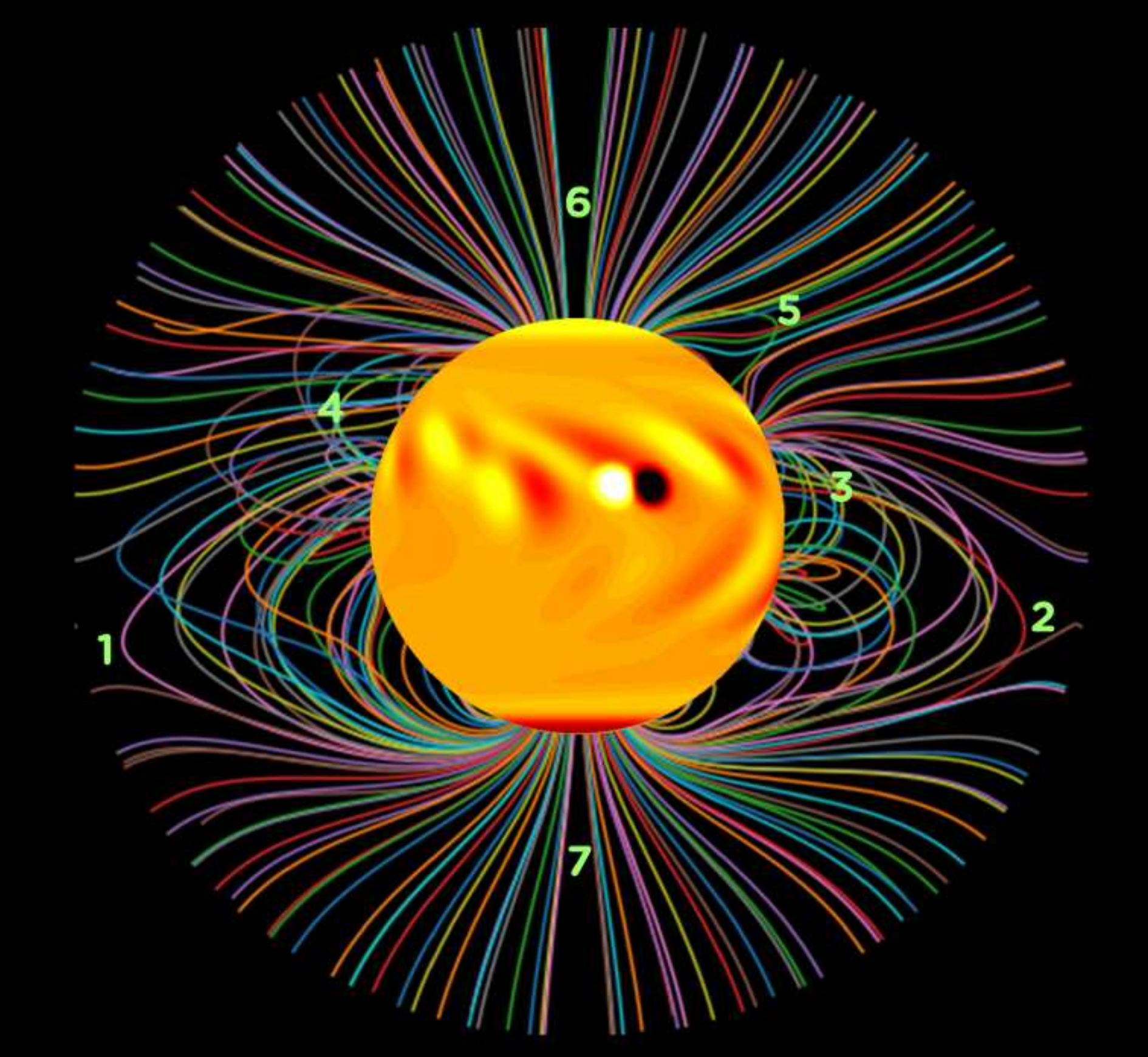}{0.45\textwidth}{(a)}
          \fig{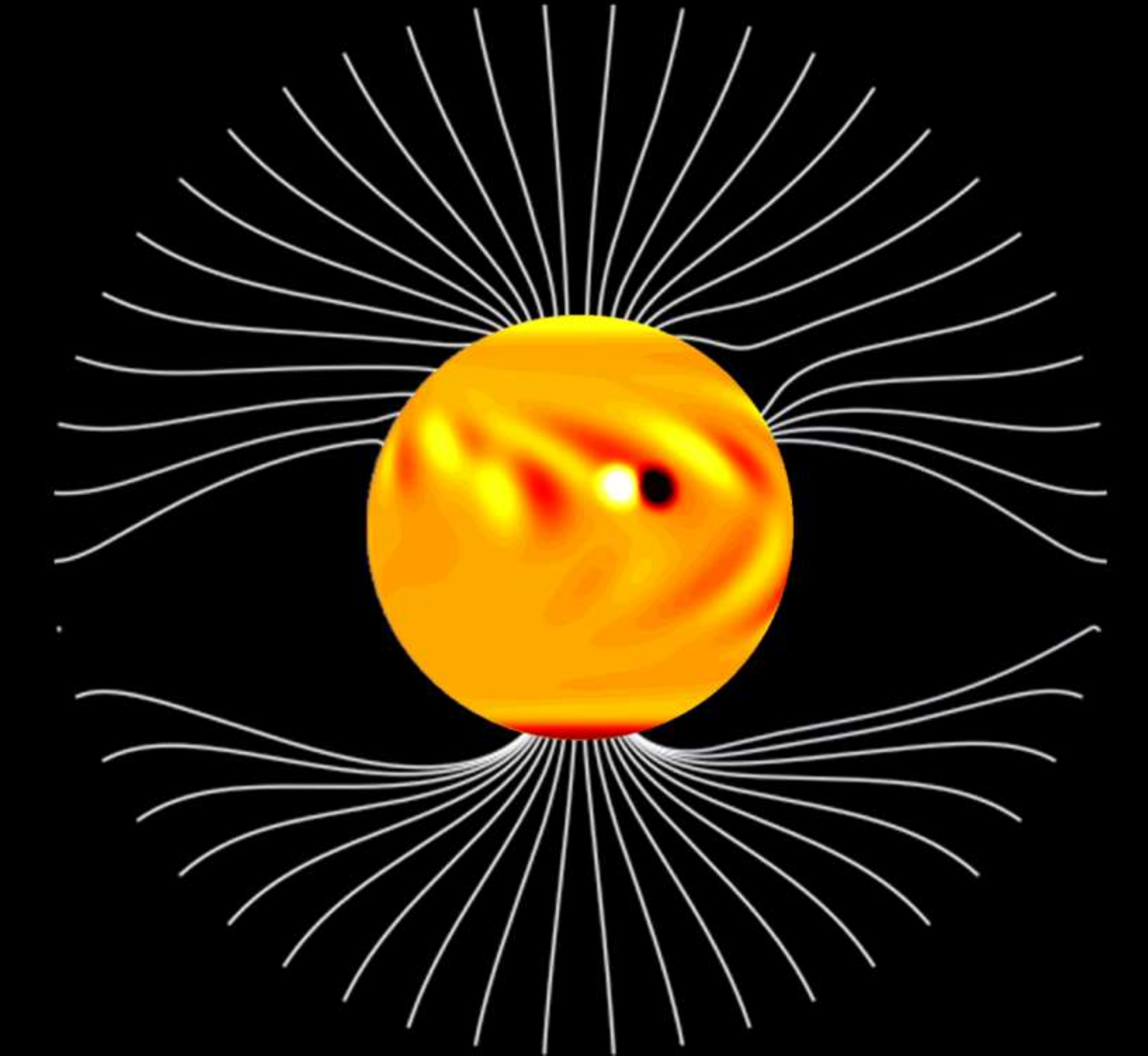}{0.45\textwidth}{(b)}}
\gridline{\fig{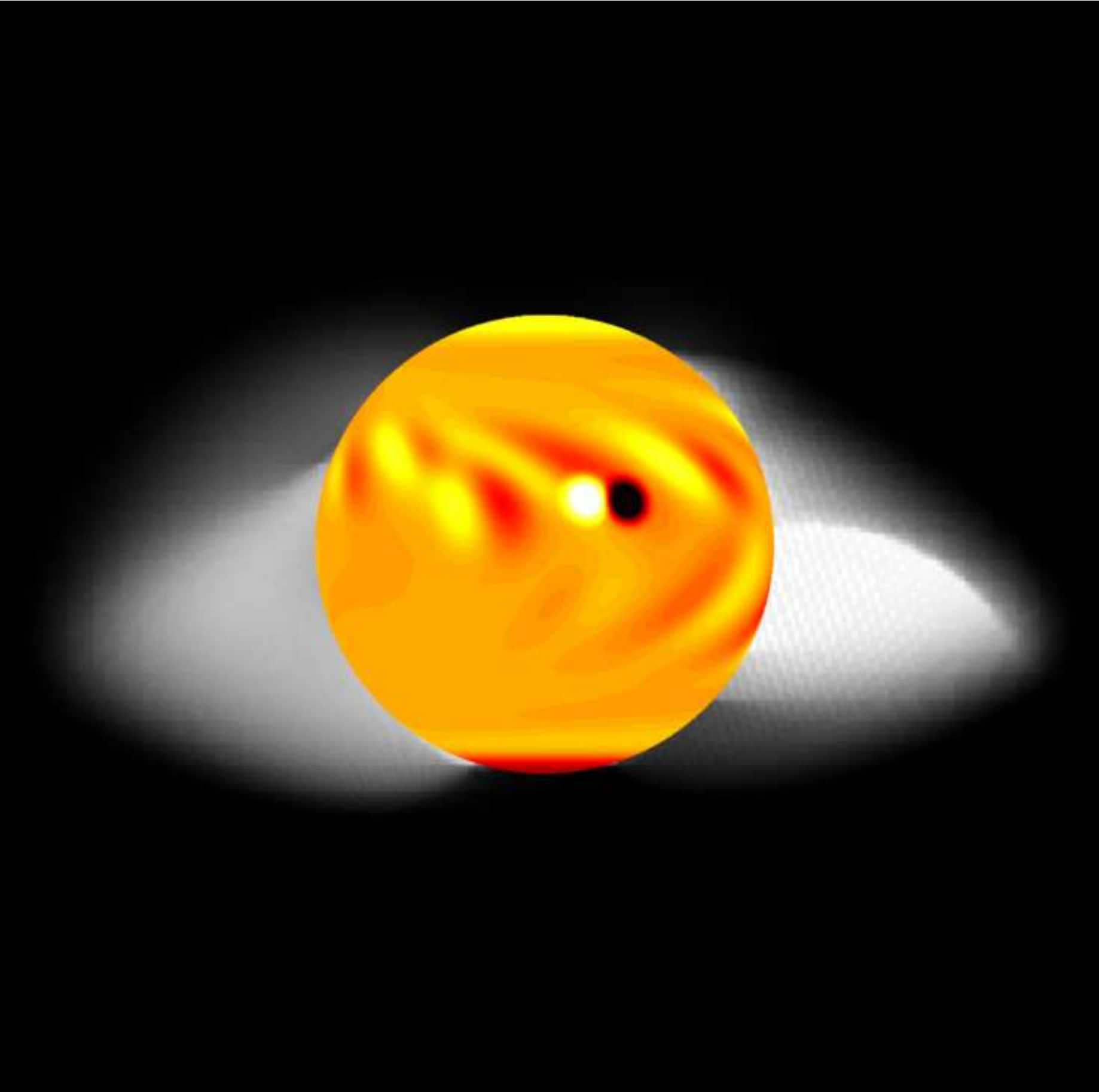}{0.45\textwidth}{(c)}
          \fig{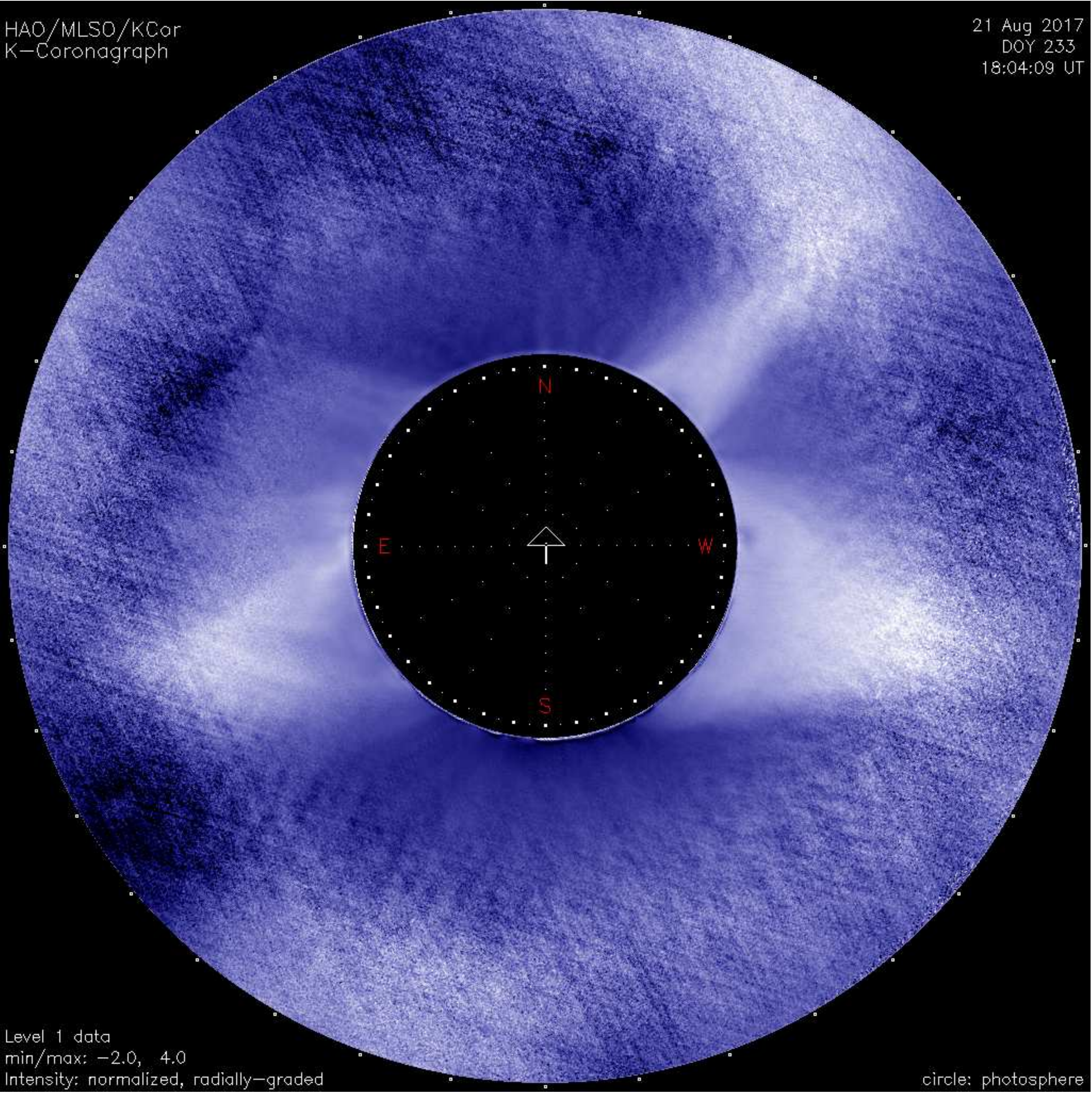}{0.45\textwidth}{(d)}}
\caption{(a) This image depicts a rendering of the open and closed solar coronal magnetic field lines generated using a PFSS model which utilized the predicted surface magnetic field map from a SFT model forward run to 21 August 2017, i.e., the day of the great American eclipse. (b): Only the open magnetic field lines reaching up to the source surface at 2.5 $R_{\odot}$ are depicted. (c): Shows a synthetic map of the white light corona expected during the eclipse rendered using a simplistic algorithm. Thus, this should be interpreted only as a guide-to-the-eye for the latitudinal location of the brightest large-scale coronal structures. (d) K-Cor coronagraph image from Mauna Loa Solar Observatory taken on the day of the solar eclipse. In all images, solar North is up.}
\end{figure}

\begin{figure}[htb!]
\gridline{\fig{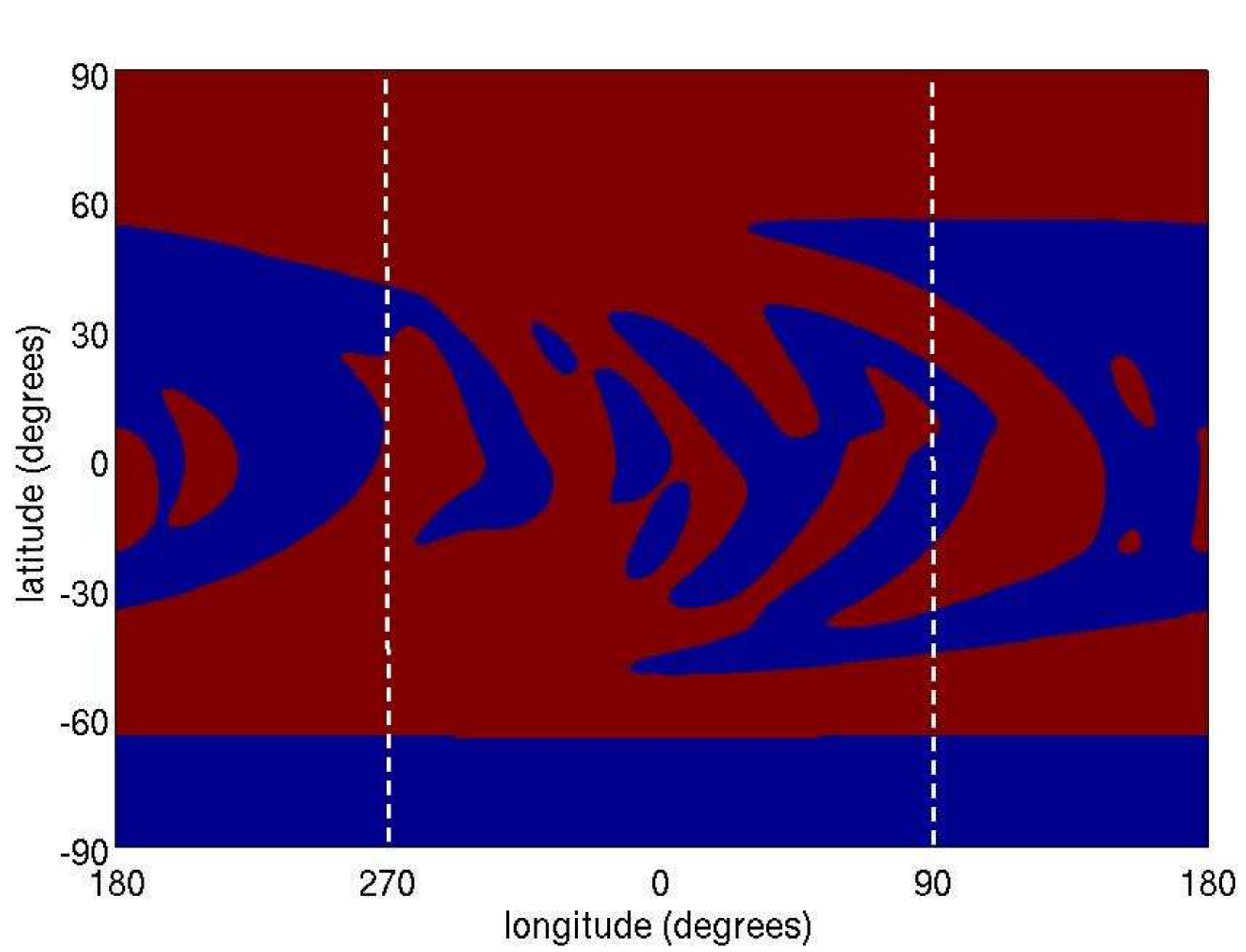}{0.32\textwidth}{(a)}
          \fig{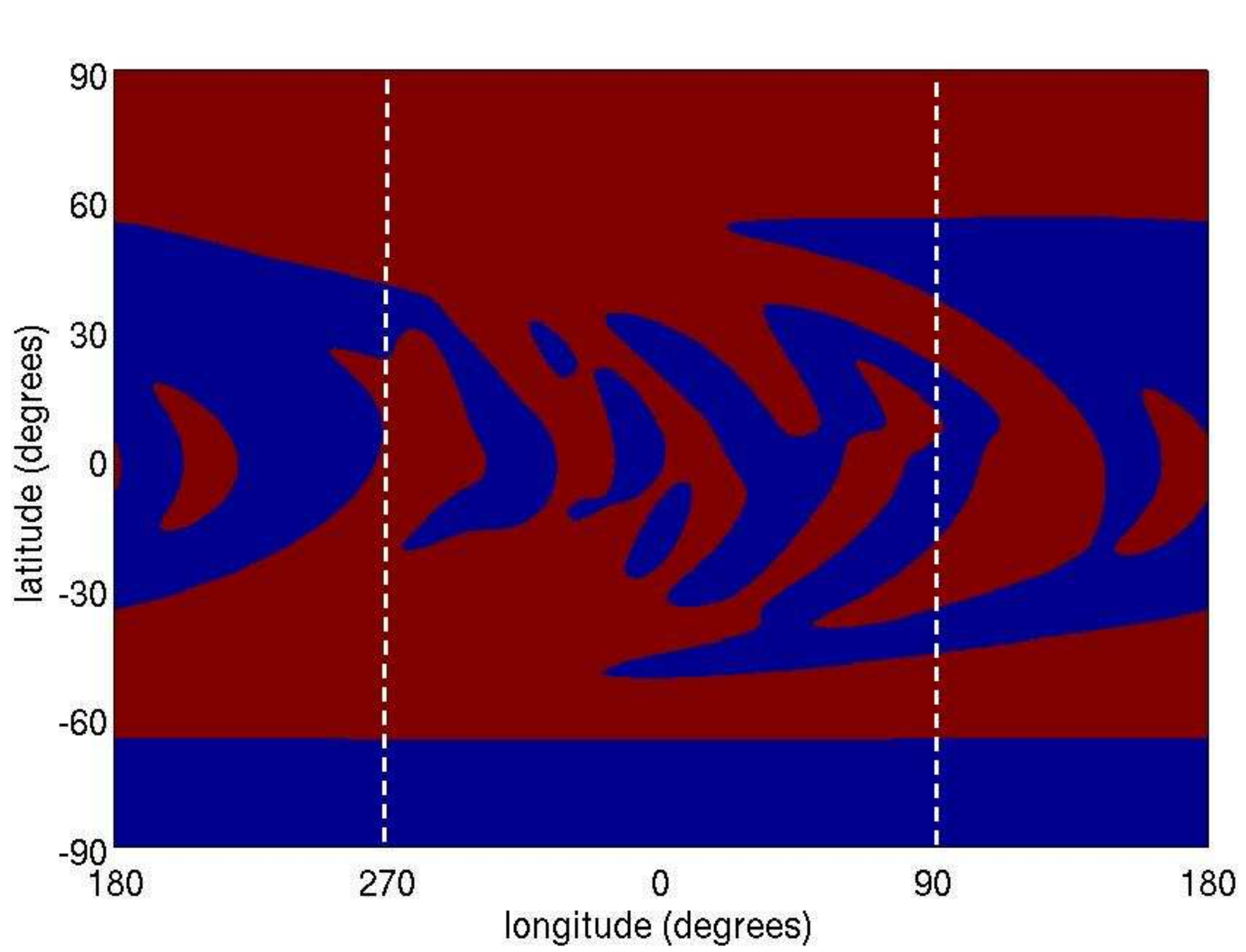}{0.32\textwidth}{(b)}
          \fig{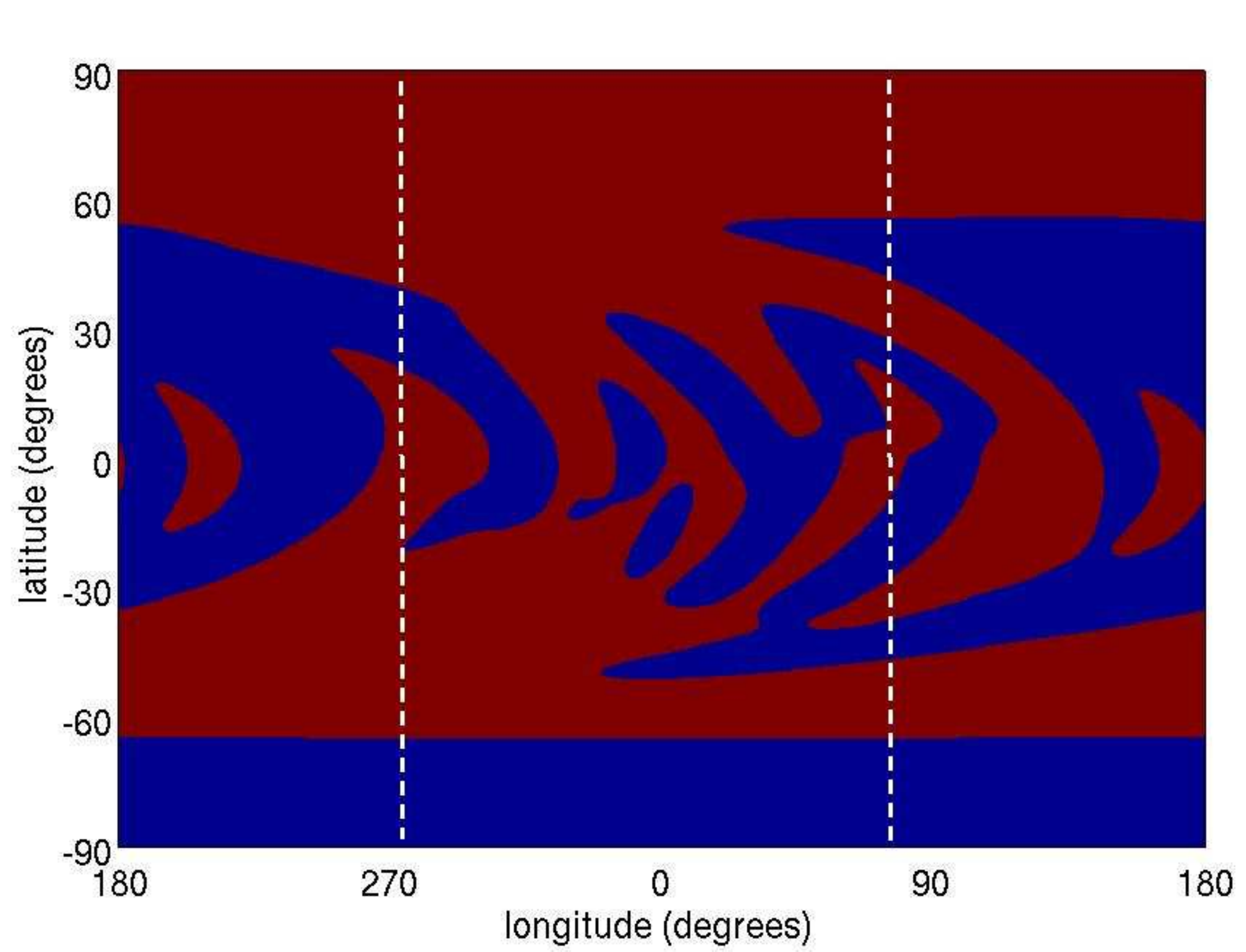}{0.32\textwidth}{(c)}}
\gridline{\fig{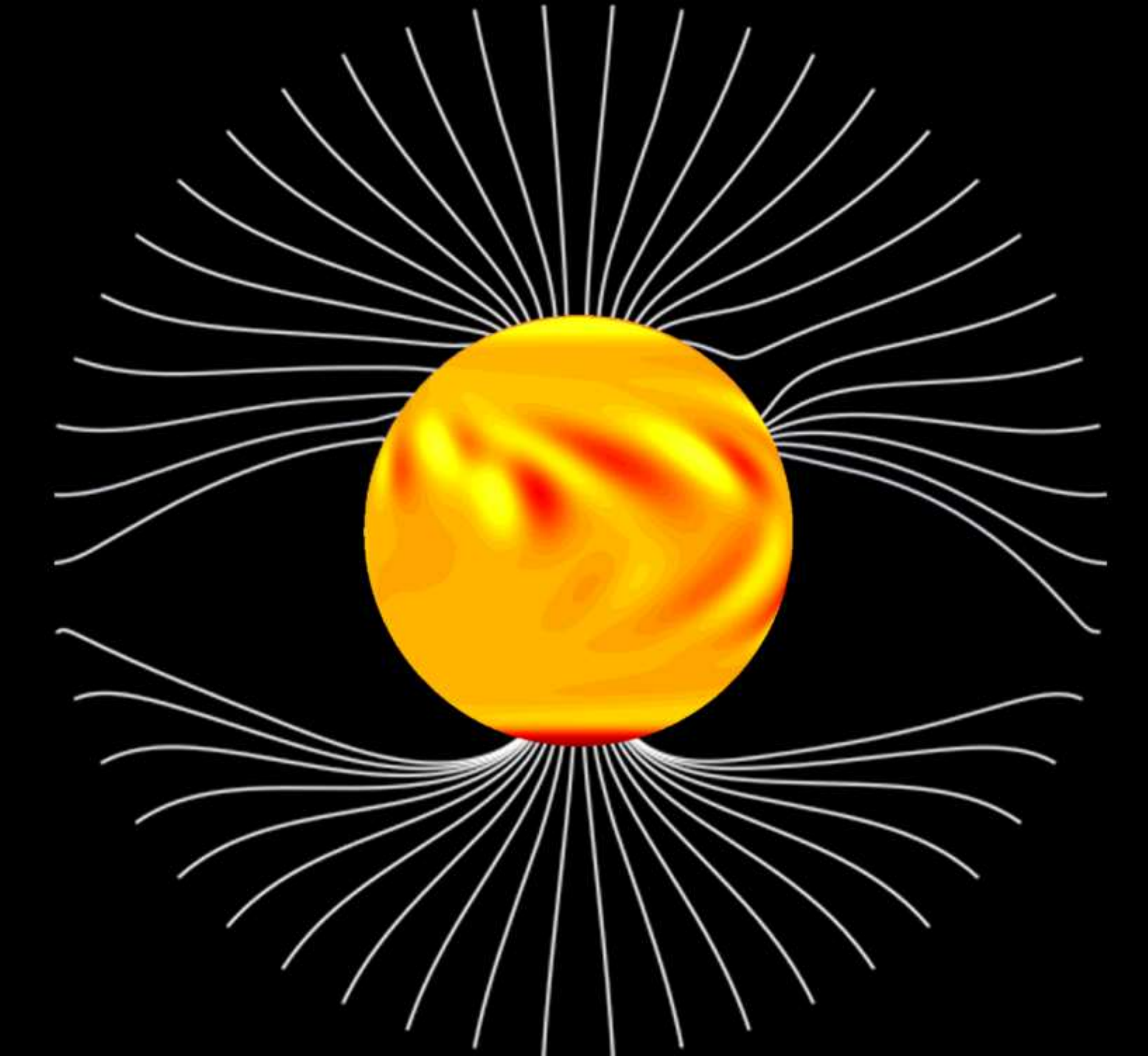}{0.3\textwidth}{(d)}
          \fig{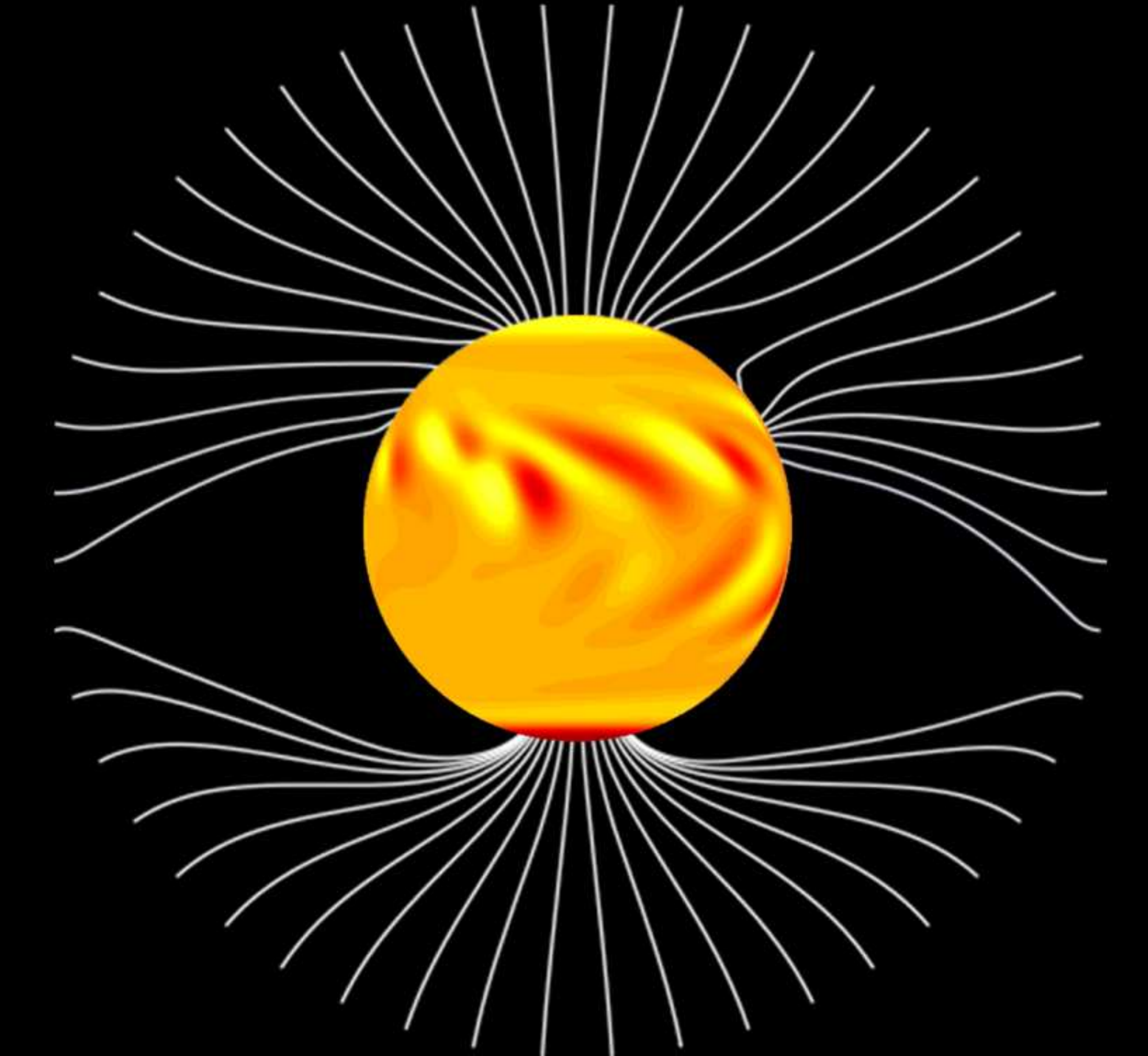}{0.3\textwidth}{(e)}
          \fig{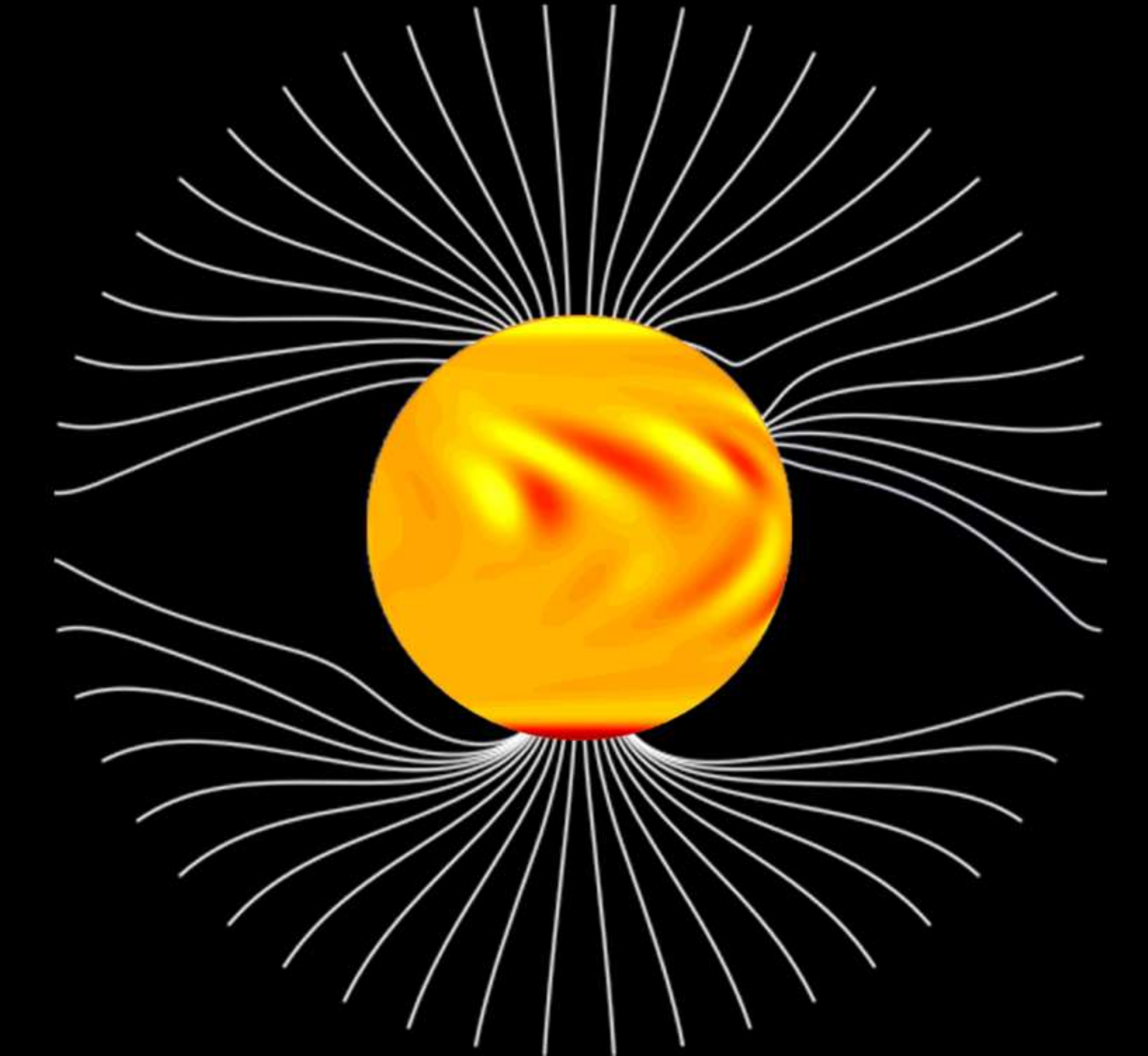}{0.3\textwidth}{(f)}}
\caption{Images in the top panel represent the polarity distribution of the SFT model predicted surface magnetic field distributions (Carrington maps) for 21 August 2017. The images are color saturated to delineate the locations of neutral lines separating regions of opposite polarity on the Sun's surface (red: positive polarity, blue: negative polarity). (a): For this prediction run, the last active region input in the SFT simulation is 24 days prior to 21 August, i.e., the duration of emergence-free evolution of the surface magnetic field is 24 days. Similarly, for (b) and (c) prediction runs, the time window of emergence-free evolution are 54 days and 76 days respectively. The set of two vertical white (dashed) lines at 270$^{\circ}$ and 90$^{\circ}$ indicate the East (Left) and West (Right) limbs of the Sun, respectively. Images in the bottom panel (d), (e) and (f) depict coronal open field line structures generated using PFSS extrapolations corresponding to simulations (a), (b) and (c), respectively.}
\end{figure}

\end{document}